\documentclass[12pt]{article}
\usepackage{amsfonts,amsmath,amssymb}
\usepackage{graphicx}
\usepackage{epic,eepic,color,graphicx}

\newfont{\twelvemsb}{msbm10 scaled\magstep1}
\newfont{\eightmsb}{msbm8}
\newfam\msbfam
\textfont\msbfam=\twelvemsb \scriptfont\msbfam=\eightmsb
\catcode`\@=11
\def\Bbb{\ifmmode\let\next\Bbb@\else
\def\next{\errmessage{Use \string\Bbb\space only in math mode}}\fi\next}
\def\Bbb@#1{{\fam\msbfam{{#1}}}}
\newcommand{\be}{\begin{equation}}
\newcommand{\ee}{\end{equation}}
\newcommand{\ba}{\begin{eqnarray}}
\newcommand{\ea}{\end{eqnarray}}

\begin{document}

\sloppy
\renewcommand{\thefootnote}{\fnsymbol{footnote}}
\newpage
\setcounter{page}{1} \vspace{0.7cm}
\begin{flushright}
28/05/08
\end{flushright}
\vspace*{1cm}
\begin{center}
{\bf The integrable $O(6)$ model and the correspondence: checks and predictions}\\
\vspace{1.8cm} {\large Francesco Buccheri \footnote{{\it Laurea} student of the University of Bologna "Alma Mater Studiorum".} 
and Davide Fioravanti
\footnote{E-mail: bucca.virus@tin.it and fioravanti@bo.infn.it .}}\\
\vspace{.5cm} {\em Sezione INFN di Bologna, Dipartimento di Fisica, Universit\`a di Bologna, \\
Via Irnerio 46, Bologna, Italy} \\
\end{center}
\renewcommand{\thefootnote}{\arabic{footnote}}
\setcounter{footnote}{0}

\begin{abstract}
{\noindent We exactly compute the energy density of the integrable $O(n)$ non-linear 
sigma model as a convergent series. 
This series is specifically analysed for the very important $O(6)$ symmetry, since it was suggested to 
result as a peculiar limit of the AdS string theory by Alday and Maldacena \cite{AM}. In this respect, 
the $O(6)$ model gives also refined confirmations and predictions once compared with the SYM Bethe 
Ansatz \cite{FGR1, FGR2}.}
\end{abstract}

%\vspace{4cm}
%{\noindent {\it Keywords}}: Integrability; counting function;
%non-linear integral equation; SYM theories. \\

\newpage

\section{The {\it questio}}

The planar $sl(2)$ sector of ${\cal N}=4$ SYM contains local composite
operators of the form
\begin{equation}
{\mbox {Tr}} ({\cal D}^s {\cal Z}^L)+.... \, , \label {sl2op}
\end{equation}
where ${\cal D}$ is the (symmetrised, traceless) covariant
derivative acting in all possible ways on the $L$ bosonic fields
${\cal Z}$. The spin of these operators is $s$ and $L$ is the
so-called 'twist'. Moreover, this sector would be described -- via
the AdS/CFT correspondence \cite {MWGKP} -- by rotating string states on the
$\text{AdS}_5\times\text{S}^5$ spacetime with $\text{AdS}_5$ and
$\text{S}^5$ charges $s$ and $L$, respectively \cite{GKP2, FT}. Proper 
superpositions of the operators (\ref{sl2op}) have definite
conformal dimension $\Delta$ depending on `t Hooft coupling $\lambda = 8 \pi^2 g^2$
\begin{equation}
\Delta = L+s+\gamma (g,s,L) \, , \label{Delta}
\end{equation}
with $\gamma (g,s,L)$ the anomalous part. In fact, the correspondence would assign 
to this dimension $\Delta$ the exact value of the energy density of a rotating string, 
provided the 't Hooft coupling is identified with the string 
tension: $\sqrt{\lambda}=\frac{R^2}{\alpha^{\prime}}$. This is indeed a duality relation 
between the coupling constants involving the semi-classical expansion on the string 
side \cite{GKP2, FT}.

A great boost in the evaluation of the anomalous $\gamma (g,s,L)$ has come from the
discovery of integrability and thus of a bethe Ansatz, although in another sector, 
the purely bosonic $so(6)$, and at one loop of the gauge theory\cite {MZ}. 
On the other hand, in the twist sector of one loop QCD the (integrable) Bethe Ansatz problem 
was at hand \cite{LIP, BDM} and later on it has been realised to be equivalent to
its supersymmetric relative with the occurence of the integrability 
extended to the whole theory and at all loops \cite{BES}. This is the scenario 
on the side of the SYM theory in the sense that, for
instance, any operator of the form (\ref{sl2op}) is associated to
one solution of some ({\it asymptotic} \footnote{This important limitation emerged 
because the Bethe Ansatz is realised by somehow using the {\it on-shell}
S-matrix \cite{Hubbard}.}) Bethe Ansatz-like equations and then any
anomalous dimension is expressed in terms of this solution. 

As a confirmation of the correspondence, integrability has been also uncovered 
and studied in the superstring theory \cite{BPR} and in this respect our interest 
will be limited to the semi-classical calculations. In this approach, the string tension
diverges since it plays the r\^ole of the inverse of the quantum Planck constant. Therefore, 
the $\lambda\rightarrow +\infty$ limit yields a power expansion 
in $1/\sqrt{\lambda}$ \cite{FTT}. Which, in particular means, that it needs to be 
implemented before any other limit and thus endowing the semiclassical calculations with a different 
limit order with respect to the gauge theory (cf. below for more details and \cite{FGR1, FGR2}).

On both the string and the gauge theory, an important 
double scaling may be considered:
\begin{equation}
s \rightarrow \infty \, , \quad L \rightarrow \infty \, , \quad
j=\frac {L}{2\ln s}={\mbox {fixed}} \, . \label {jlimit}
\end{equation}
In fact, the relevance of this logarithmic scaling for the anomalous
dimension has been pointed out and deeply studied in \cite{FTT} and \cite{AM} within 
the semi-classical string theory (cf. also \cite{BGK} within the one-loop SYM
theory). Moreover, these long operators with large spin have been recently shown 
to satisfy the Sudakov scaling \cite{AM, FRS}
\begin{equation}
\gamma (g,s,L)=f(g,j)\ln s + O((\ln s)^{-\infty}) \, \label{Suds},
\end{equation}
which generalises the one loop result of \cite{BGK}.
\footnote{$O({\ln s}^{-\infty})$ means a remainder which goes faster
that any power of $\ln s$: $\lim\limits_{s \rightarrow \infty} (\ln
s)^k O((\ln s)^{-\infty}) =0, \forall \, k>0$.} 
Actually, in \cite{FRS} this statement was argued by computing iteratively the solution of some
integral equations and then, thereof, {\it the generalised scaling
function}, $f(g,j)$ at the first orders in $j$ and $g^2$: more
precisely the first orders in $g^2$ have been computed for the first {\it
generalised scaling functions} $f_n(g)$, forming
\begin{equation}
f(g,j)=\sum _{n=0}^{\infty} f_n(g)j^n \, .
\end{equation}
As a by-product, the reasonable conjecture has been put forward that
the two-variable function $f(g,j)$ should be analytic (in
$g$ for fixed $j$ and in $j$ for fixed $g$). In \cite{BFR2} similar
results have been derived for what concerns the contribution beyond
the leading scaling function $f(g)=f_0(g)$, but with a modification which has
allowed to neglect the non-linearity for finite $L$ and to end-up with one
linear integral equation (LIE). The latter does not differ from the
BES one (which covers the case $j=0$, cf. the last of \cite{BES}), but for the
inhomogeneous term. Moreover, a suitable modification of this LIE has been applied 
in \cite{FGR1, FGR2} to derive still a LIE in the scaling (\ref{jlimit}) 
(for any $g$ and $j$). This is indeed, a way to determine the generalised scaling function
$f(g,j)$ and also its {\it constituents} $f_n(g)$ for all values of $j$ and $g$, 
thus interpolating from weak to strong coupling. Today, an interesting paper \cite{BK} appears which seems
to have some equivalent equation, coming from \cite{FRS}, from which it apparently derives a map to 
the $O(6)$ sigma model and the leading strong coupling 
behaviour of $f_3$ (as $f_2=0$ appeared already in \cite{FRS} and $f_1$ in \cite{FGR1}).  

In the following, we will constrain ourselves to the analysis of the $O(N)$ energy density of the 
non linear sigma models, since one representative, the $O(6)$ model, was suggested by Alday and Maldacena 
\cite{AM} to represent the limit theory with small $SO(6)$ charge $j\gg \sqrt{\lambda}$. For in 
this regime the masses of the fermions do not contribute to the energy density, but to $f_0(g)$, where 
they give the natural UV cut-off. Therefore, we will show that the additional energy density, 
$2\Omega(g,j)$ (cf. (\ref{adden})) contained in $f(g,j)$ can be computed exactly at least at the leading 
mass gap ($m$ of \ref{mass}) order by means of the $O(6)$ computations. Hence, we will compute below the $O(6)$ 
energy density as a convergent series in $j/m$, 
checking a perfect agreement with the gauge theory computations of \cite{FGR2} up to the first 
interaction and model depending term $f_4(g)$, i.e. $\Omega_4$. An all order 
explicit match would be desirable for the future and in this respect the following $O(6)$ model results give 
a series of exact predictions.

\section{The O(N) nonlinear sigma model}
We want to show that the result in \cite{HMN} for the energy density $\Omega$ as a function of the particle density $j$ in the strong coupling regime ($g>>1$) with $j<<m$, namely the ``non-interacting fermion gas'' approximation, keeps correct when all orders in B are retained in the calculation, i.e. that it is the correct result up to the third order $j^3$. In formulas, that
\begin{equation} \label{fermiongas}
  \Omega=m^2\left( \frac{j}{m}+\frac{\pi^2}{6} {\left( \frac{j}{m}\right) }^3 + O\left( \left( \frac{j}{m}\right) ^4\right) \right) 
\end{equation}
where the mass gap is (see \cite{BZ} and \cite{AM}, with $\sqrt{\lambda}=\frac{1}{t}$)
\begin{equation} \label{mass}
  m=\frac{2^{3/4}\pi^{1/4}}{\Gamma(5/4)}g^{1/4}e^{-\pi g}+\ldots
\end{equation}

We begin in full generality, by considering a O(N) nonlinear sigma model. A manipulation of BA equations \cite{HMN} provides a "pseudoenergy" $ \varepsilon(\vartheta) $ as the solution of a linear integral equation of Fredholm type
\begin{displaymath} \label{epsiloneqn}
 \varepsilon(\vartheta)-\int_{-B}^{B}K(\vartheta-\vartheta')\varepsilon(\vartheta')d\vartheta'=h-m\cosh\vartheta
\end{displaymath}
With the boundary condition 
\begin{equation} \label{bound}
\varepsilon(B)=0
\end{equation}
 The condition (\ref{bound}) allows us, at least in principle, to determine the value of the parameter $B$.
The kernel $K\left( \vartheta\right) $ comes from the \cite{ZamS} two-particle S-matrix\footnote{This matrix contains an ambiguity of the CDD kind \cite{CDD}, as underlined by Zamolodchikov and Zamolodchikov. What people generally do, is to use the most simple solution (\ref{Smatrix})}
\begin{equation} \label{Smatrix}
 S\left( \vartheta\right) = - \frac{\Gamma\left(1+\frac{i\vartheta}{2\pi} \right) \Gamma\left(\frac{1}{2}-\frac{i\vartheta}{2\pi}\right) \Gamma\left(\frac{1}{2}+\Delta+ \frac{i\vartheta}{2\pi} \right) \Gamma\left(\Delta- \frac{i\vartheta}{2\pi} \right) } {\Gamma\left(1-\frac{i\vartheta}{2\pi} \right) \Gamma\left(\frac{1}{2}+\frac{i\vartheta}{2\pi}\right) \Gamma\left(\frac{1}{2}+\Delta - \frac{i\vartheta}{2\pi} \right) \Gamma\left(\Delta+ \frac{i\vartheta}{2\pi} \right)}
\end{equation}
through
\begin{equation} \label{kerneldef}
 K\left( \vartheta\right)=\frac{1}{2\pi i}\frac{\partial}{\partial \vartheta}\log S\left( \vartheta\right) 
\end{equation}
where $\Delta=\frac{1}{N-2}$ What is more, the Bethe Ansatz procedure gives the Helmholtz free energy $f$ as
\begin{equation} \label{helmholtz}
-\frac{2\pi}{m} f\left( h,m\right) = {\displaystyle\intop_{-B}^{B} \cosh\vartheta \varepsilon \left( \vartheta\right)d\vartheta}
\end{equation}
while general Thermodynamics provides the density of particles
\begin{equation} \label{density j}
j=-\frac{\partial}{\partial h}f(h,m)
\end{equation}
Gathering up all the elments, an optimistic procedure would be that of solving (\ref{epsiloneqn}) for $\epsilon$, put it into (\ref{helmholtz}), then calculate $j$ by (\ref{density j}) and, finally, $\Omega$ by Legendre transform
\begin{equation}  \label{legendre}
 \Omega\left( j,m \right) = f\left( h,m\right) + jh
\end{equation}
The ``non-interacting fermion gas'' approximation corresponds to the limit $B\rightarrow 0$ in (\ref{epsiloneqn}), i.e. $\frac{h-m}{m}\rightarrow 0^+$. Indeed, considering $S(\vartheta)= -1$ and $K(\vartheta)=\frac{1}{S(\vartheta)}\frac{d}{d\vartheta}S(\vartheta)=0$, is equvalent to saying that $B\rightarrow 0$ allows one to retain only the forcing therm of (\ref{epsiloneqn}), that is the zero-th order approximation of the Liouville-Neumann procedure.

%%%%%%%%%%%%%%%%%%%%%%%%%%%%%%%%%%%%%%%%%%%%%%%%%%%%%%%%%%%%%%%%%%%%%%%%%%%%%%%%%%%%%%%%%%%%%%%%%%%%%%%%%%%%%%%%%%%%%%%%%%%%%%
\section{The 1D non-interacting fermion gas}
Despite its simplicity, this model is noteworthy because it represents a paradigmatic example of how Bethe Ansatz works and also because it produces results that can directly be compared with those coming from more advanced calculations from the SYM$_4$ front.  What we will call "fermions" are introduced in a simple fashion by defyining their two-particle S-matrix as
\begin{equation}
S(\vartheta)=-1 
\end{equation}
now we can set in motion the calculation sketched above, with $K(\vartheta)=0$ for (\ref{kerneldef}). Then, by (\ref{epsiloneqn}),(\ref{density j}), (\ref{helmholtz}), (\ref{legendre}) we have
\begin{equation}
\varepsilon(\vartheta)=h-m\cosh\vartheta
\end{equation}
\begin{equation}
f(h)-f(0)=-\frac{m}{2\pi}\intop_{-B}^B(h-m\cosh\vartheta)\cosh\vartheta d\vartheta =-\frac{m}{2\pi}\left[ 2h\sinh B -mB -m\cosh B\sinh B\right]
\end{equation}
\begin{equation}
j=\frac{m}{2\pi}\intop_{-B}^B\cosh\vartheta d\vartheta =\frac{m}{\pi}\sinh B \qquad\qquad\qquad B=\arcsin h \left(\frac{\pi}{m}j\right)
\end{equation}
\begin{eqnarray}
\Omega&=&f+jh=\frac{m^2}{2\pi}\left[B+\sinh B \cosh B \right] =\frac{m^2}{2\pi}\left[\arcsin h\left(\frac{\pi}{m}j\right)+\frac{\pi}{m}j\sqrt{1+\left(\frac{\pi}{m}j\right)^2} \right] = \nonumber \\
&=&\frac{m^2}{2\pi} \left[ \frac{\pi}{m}j -\frac{1}{6}\left(\frac{\pi}{m}j\right)^3 +\frac{3}{40}\left(\frac{\pi}{m}j\right)^5 +
   \frac{\pi}{m}j \left( 1 + \frac{1}{2}\left(\frac{\pi}{m}j\right)^2 - \frac{1}{8}\left(\frac{\pi}{m}j\right)^4 \right)+ O\left(\frac{j^7}{m^7}\right)\right]  \nonumber \\
&=&\frac{m^2}{\pi}\left[\frac{\pi}{m}j + \frac{1}{6}\left(\frac{\pi}{m}j\right)^3 - \frac{1}{40}\left(\frac{\pi}{m}j\right)^5+O\left(\frac{j^7}{m^7}\right)\right]
\end{eqnarray}
The series contains only the odd powers of $j$: we will see that turning on an interaction will turn on the coefficients of the even,  powers from the fourth power on, together with affecting the odd terms. 

\section{The relativistic interacting gas: some preliminaries}
 The formal solution of (\ref{epsiloneqn}) is easily seen to be
\begin{eqnarray}\label{solution}
\varepsilon(\vartheta)&=&h-m\cosh\vartheta+\int_{-B}^{B}K(\vartheta-\vartheta')\left(h-m\cosh\vartheta'+\int_{-B}^{B}K(\vartheta'-\vartheta'')\varepsilon(\vartheta'')d\vartheta''\right)d\vartheta'=\nonumber \\ 
&=&\ldots=  h-m\cosh\vartheta+ \nonumber \\ &+&{\displaystyle\sum_{n=1}^{\infty}}\intop_{-B}^{B}d\vartheta_{1}...\intop_{-B}^{B}d\vartheta_{n}K(\vartheta-\vartheta_{1})K(\vartheta_{1}-\vartheta_{2})...K(\vartheta_{n-1}-\vartheta_{n})\left(h-m\cosh\vartheta_{n}\right)=    \\
&=&1+{\displaystyle\sum_{n=1}^{\infty}}B^{n}\intop_{-1}^{1}dx_{1}...\intop_{-1}^{1}dx_{n}K(Bx-Bx_{1})K(Bx_{1}-Bx_{2})  \nonumber  \\
&& \qquad\qquad\qquad\qquad\qquad\qquad\qquad\qquad\ldots K(Bx_{n-1}-Bx_{n})\left(h-m\cosh(Bx_{n})\right)
\end{eqnarray}
where
\begin{eqnarray}
K\left(\vartheta\right)&=&\frac{1}{4\pi^2}\Big[\psi\left(1+\frac{i\vartheta}{2\pi}\right)-\psi\left(\frac{1}{2}+\frac{i\vartheta}{2\pi}\right)+\psi\left(\frac{1}{2}+\Delta+\frac{i\vartheta}{2\pi}\right)-\psi\left(\Delta+\frac{i\vartheta}{2\pi}\right) +\nonumber \\
&+&\psi\left(1-\frac{i\vartheta}{2\pi}\right)-\psi\left(\frac{1}{2}-\frac{i\vartheta}{2\pi}\right)+\psi\left(\frac{1}{2}+\Delta-\frac{i\vartheta}{2\pi}\right)-\psi\left(\Delta-\frac{i\vartheta}{2\pi}\right)\Big]
\end{eqnarray}
with
\begin{displaymath}
 \psi\left(x\right)=\frac{\Gamma'\left(x\right)}{\Gamma\left(x\right)}=\frac{d}{dx}ln\Gamma\left(x\right)=-\gamma+{\displaystyle \sum_{n=0}^{\infty}}\left(\frac{1}{n+1}-\frac{1}{n+x}\right)
\end{displaymath}
where $\gamma$ is the Euler-Mascheroni constant. Its $h$-derivative, by the aid of (\ref{epsiloneqn}) and (\ref{bound}) reads
\begin{eqnarray} \label{h-derivative}
\frac{\partial\varepsilon_h(\vartheta)}{\partial h} &=&1 + \left( K\left( \vartheta-B\right)\varepsilon\left( B\right) \right)\frac{\partial B}{\partial h}  + \intop_{-B}^{B}d\vartheta K\left( \vartheta-\vartheta'\right)\frac{\partial\varepsilon_h(\vartheta')}{\partial h}   \nonumber \\
&=&1 + \intop_{-B}^{B}d\vartheta
        K\left(\vartheta-\vartheta'\right)\frac{\partial\varepsilon_h(\vartheta')}{\partial h}
\end{eqnarray}
that is, another Fredholm integral equation, with a constant forcing therm. Its solution is
\begin{eqnarray}\label{hderivative}
\frac{\partial\varepsilon_h(\vartheta)}{\partial h}&=&1+{\displaystyle\sum_{n=1}^{\infty}}\intop_{-B}^{B}d\vartheta_{1}...\intop_{-B}^{B}d\vartheta_{n}K(\vartheta-\vartheta_{1})K(\vartheta_{1}-\vartheta_{2})...K(\vartheta_{n-1}-\vartheta_{n}) =\nonumber  \\
&=&1+{\displaystyle\sum_{n=1}^{\infty}}B^{n}\intop_{-1}^{1}dx_{1}...\intop_{-1}^{1}dx_{n}K(Bx-Bx_{1})K(Bx_{1}-Bx_{2})\nonumber  \\
&& \qquad\qquad\qquad\qquad\qquad\qquad\qquad\qquad\ldots K(Bx_{n-1}-Bx_{n})
\end{eqnarray}
By the use of the formula
\begin{displaymath}
 \psi\left(\frac{1}{2}+\frac{1}{2}x\right)-\psi\left(\frac{1}{2}x\right)=2{\displaystyle \sum_{k=0}^{\infty}}\frac{\left(-1\right)^{k}}{k+x}
\end{displaymath}
we can rewrite the Kernel as
\begin{eqnarray}
K(x)&=&\frac{1}{2\pi}\Big[{\displaystyle \sum_{k=0}^{\infty}}\frac{\left(-1\right)^{k}}{\pi(k+1)+i(\vartheta-\vartheta')}+{\displaystyle \sum_{k=0}^{\infty}}\frac{\left(-1\right)^{k}} {\pi(k+2\Delta)+i(\vartheta-\vartheta')}+		\nonumber  \\
&+&{\displaystyle\sum_{k=0}^{\infty}}\frac{\left(-1\right)^k}{\pi\left(k+1\right)-i\left(\vartheta-\vartheta'\right)}+{\displaystyle\sum_{k=0}^{\infty}}\frac{\left(-1\right)^{k}}{\pi\left(k+2\Delta\right)-i\left( \vartheta-\vartheta'\right) }\Big]=  \nonumber \\ &=&\frac{1}{2\pi}\left[{\displaystyle\sum_{k=0}^{\infty}}\frac{\left(-1\right)^{k}2\pi(k+1)}{\pi^{2}(k+1)^{2}+(\vartheta-\vartheta')^{2}}+{\displaystyle\sum_{k=0}^{\infty}}\frac{\left(-1\right)^{k}2\pi(k+1)}{\pi^{2}(k+2\Delta)^{2}+(\vartheta-\vartheta')^{2}}\right]		\nonumber \\  &=&\frac{1}{\pi^{2}}{\displaystyle\sum_{k=0}^{\infty}}\left[\frac{1}{k+1}\frac{\left(-1\right)^{k}}{1+\left(\frac{\vartheta-\vartheta'}{\pi(k+1)}\right)^{2}}+\frac{1}{k+2\Delta}\frac{\left(-1\right)^{k}}{1+\left(\frac{\vartheta-\vartheta'}{\pi(k+2\Delta)}\right)^{2}}\right]
\end{eqnarray}
Some comments are to be made. First of all, even powers of $j$ do not appear in this expansion, because the energy density is written as the sum of two odd function. Of course, it follows that all even powers in the interacting case will vanish when we shut down the interaction. What is more, the first three term are somehow "stable" (see comments to (\ref{Omegarapida}) against the insertion of a less trivial S-matrix.

%%%%%%%%%%%%%%%%%%%%%%%%%%%%%%%%%%%%%%%%%%%%%%%%%%%%%%%%%%%%%%%%%%%%%%%%%%%%%%%%%%%%%%%%%%%%%%%%%%%%%%%%%%%%%%%%%%%%%%%%%%%%%%%%%%%%

\section{The very first orders in $B$}
We recall that in the O(6) strong-coupling regime we have $B<<1$. This way written, the kernel allows a easier calculation of the first few orders in B of the quantity $\varepsilon(\vartheta)$.
The first, raw approximation consists in writing the solution $\varepsilon(\vartheta)$ as composed of the forcing therm alone, ${\varepsilon(\vartheta)=h-m\cosh\vartheta}$
The boundary condition implies
\begin{equation}
 {B=arccosh\left( \frac{h}{m}\right) \simeq \sqrt{2\frac{h-m}{m}}}
\end{equation}
This bare result is already enough to obtain the fermion gas approximation, as already stated.
We would like to know how much this result for the energy density is reliable, i.e. if calculations involving higher orders in B are likely to leave the first coefficient as they are, or else if they are destined to upset the solution. We will go on by brute-force Taylor expansion in B. To begin, we illustrate the second order case, an then we will give the systematics of the method.

\subsection{What we can do up to the order $B^2$}
By defining the new variable $x=\frac{\vartheta}{B}$ and expanding the fractions that compose the kernel in $B$ we obtain, up to the order $B^{2}$
\begin{eqnarray}
\varepsilon(\vartheta)&=&h-m\cosh\vartheta+B\int_{-1}^{1}K(\vartheta-Bx)\varepsilon(Bx)dx=  \nonumber \\
&=&h-m\cosh\vartheta+B\int_{-1}^{1}K(\vartheta-Bx)\left(h-m\cosh Bx+B\int_{-1}^{1}K(Bx-By)\varepsilon(By)dy\right)dx= \nonumber \\ &=&h-m\cosh\vartheta+\frac{B}{\pi^{2}}\int_{-1}^{1} dx{\displaystyle \sum_{k=0}^{\infty}}\left[\frac{1}{k+1}\frac{\left(-1\right)^{k}}{1+\left(\frac{\vartheta-Bx}{\pi(k+1)}\right)^{2}}+\frac{1}{k+2\Delta}\frac{\left(-1\right)^{k}}{1+\left(\frac{\vartheta-Bx}{\pi(k+2\Delta)}\right)^{2}}\right] \nonumber \\
& & \left( h-m\cosh Bx +\frac{B}{\pi^{2}}\int_{-1}^{1}{\displaystyle\sum_{k=0}^{\infty}}\left[\frac{1}{k+1}\frac{\left(-1\right)^{k}}{1+\left(\frac{Bx-By} {\pi(k+1)}\right)^{2}}+\frac{1}{k+2\Delta}\frac{\left(-1\right)^{k}}{1+\left(\frac{Bx-By}{\pi(k+2\Delta)}\right)^{2}}\right]\varepsilon(By)dy\right)dx=  \nonumber \\
&=&h-m\cosh\vartheta+\frac{B}{\pi^{2}}{\displaystyle \sum_{k=0}^{\infty}}\int_{-1}^{1}\left[\frac{\left(-1\right)^{k}}{k+1}\left(1-\left(\frac{\vartheta-Bx}{\pi(k+1)}\right)^{2}\right)+\frac{\left(-1\right)^{k}}{k+2\Delta}\left(1-\left(\frac{\vartheta-Bx}{\pi(k+2\Delta)}\right)^{2}\right)\right] \nonumber \\ 
 &&\left(h-m\cosh Bx+\frac{B}{\pi^{2}}{\displaystyle \sum_{k=0}^{\infty}}\int_{-1}^{1}\left(\frac{\left(-1\right)^{k}}{k+1}+\frac{\left(-1\right)^{k}}{k+2\Delta}\right)\varepsilon(By)dy\right)dx=	\nonumber \\
&=&h-m\cosh\vartheta+\frac{B}{\pi}\int_{-1}^{1}\left[S_1-S_3\left(\vartheta-Bx\right)^{2}\right]   \nonumber \\
&&\qquad\qquad\qquad\qquad\qquad\qquad\left[h-m\cosh(Bx)+\frac{B}{\pi}\int_{-1}^{1}S_1\left(h-m\cosh By\right)dy\right]dx
\end{eqnarray}
where
\begin{displaymath}
 S_1=\frac{1}{\pi}{\displaystyle\sum_{k=0}^{\infty}}\left[\frac{\left(-1\right)^{k}}{k+1}+\frac{\left(-1\right)^{k}}{k+2\Delta}\right]
\mbox{\qquad\qquad and \qquad\qquad}
S_3={\displaystyle \frac{1}{\pi^3} \sum_{k=0}^{\infty}}\left[\frac{\left(-1\right)^{k}}{\left(k+1\right)^{3}}+\frac{\left(-1\right)^{k}}{\left(k+2\Delta\right)^{3}}\right]
\end{displaymath}
\begin{eqnarray}
\varepsilon(\vartheta)&\simeq&h-m\cosh\vartheta+ \nonumber  \\
&&+\frac{B}{\pi}\int_{-1}^{1}\left[S_1-S_3\left(\vartheta-Bx\right)^{2}\right]
  \left[h-m\left(1+\frac{1}{2}B^{2}x^{2}\right)+\frac{B}{\pi}\int_{-1}^{1}S_1\left(h-m\right)dy\right]dx=  \nonumber \\
& \simeq& h-m\cosh\vartheta+\frac{B}{\pi}\int_{-1}^{1}\left[S_1-S_3\left(\vartheta-Bx\right)^{2}\right]
    \left[\left(h-m\right)-\frac{m}{2} B^{2}x^{2}+2\frac{B}{\pi}S_1\left(h-m\right)\right]dx=   	\nonumber  \\
& \simeq & h-m\cosh\vartheta+\frac{B}{\pi}\Bigg(2S_1\left(h-m\right) -
    \frac{m}{2}S_1B^{2}\int_{-1}^{1}x^{2}dx+4\frac{B}{\pi}S_1^{2}\left(h-m\right)+                      \nonumber \\ 
& & - S_3\left(h-m\right)\int_{-1}^{1}\left(\vartheta-Bx\right)_{}^{2}dx+
    \frac{m}{2}S_3B^{2}\int_{-1}^{1}\left(\vartheta-Bx\right)_{}^{2}x^{2}dx+                       \nonumber  \\ 
& & -2\frac{B}{\pi}S_1S_3\int_{-B}^{B}\left(\vartheta-Bx\right)_{}^{2}\Bigg)dx
\end{eqnarray}
\begin{eqnarray}
\varepsilon(B)& = &h-m\left( 1+ \frac{1}{2}B^2 \right) + \frac{2}{\pi}(h-m)S_1 B+ \frac{4}{\pi^2}(h-m)S_1^2 B^2+o(B^3) \nonumber \\
&=& h-m +\frac{2}{\pi}(h-m)S_1 B+\left(-\frac{m}{2}+\frac{4}{\pi^2}(h-m)S_1^2\right) B^2+o(B^3)
\end{eqnarray}
By solving the condition $\varepsilon(B)=0$ we gain an expression for the extreme
\begin{eqnarray} \label{B2}
 B & \simeq &\frac{-(h-m)\frac{S_1}{\pi}+
   \sqrt{\frac{1}{\pi^2}S_1^2(h-m)^2+\frac{m}{2}(h-m)-\frac{4}{\pi^2}(h-m)^2S_1^2}}{\frac{4}{\pi^2}(h-m)S_1^2-
    \frac{m}{2}}= \nonumber \\
&=&\left(-(h-m)\frac{S_1}{\pi}-\sqrt{\frac{m}{2}(h-m)}\left(1-\frac{3}{2\pi^2}S_1^2(h-m)^2\right)\right)
    \frac{2}{-m}\left(1+8\frac{S_1^2}{\pi^2}\frac{h-m}{m} \right)+ \ldots = \nonumber  \\
&= & \sqrt{\frac{2}{m}(h-m)}+\frac{2}{\pi}S_1\frac{h-m}{m}+O\left( \left( \frac{h-m}{m}\right)
    ^\frac{3}{2} \right) 
\end{eqnarray}
where we have coherently omitted all contributions from $B^3$ on, that is $ \left( \frac{h-m}{m}\right) ^\frac{3}{2} $ and higher powers.
For what the free energy is concerned, we can write the BA formula and approximate to the second order in $B$
\begin{eqnarray} \label{2ndorderfreenrg}
-\frac{2\pi}{m} f\left( h,m\right) &=& {\displaystyle\intop_{-B}^{B} \cosh\vartheta \varepsilon \left( \vartheta\right)d\vartheta} = B{\displaystyle\intop_{-1}^{1} \cosh\left( B\,x_0 \right)  \varepsilon \left( B\,x_0 \right)dx_0}                                                   \nonumber \\
&\simeq&  B{\displaystyle\intop_{-1}^{1} \cosh\left( B\,x_0 \right)\left( h-m\cosh\left( B\,x_0 \right)+B\intop_1^1S_1(h-m)dx_1 \right)dx_0}                            \nonumber \\
&\simeq&  B{\displaystyle\intop_{-1}^{1} \left( 1+\frac{1}{2}B^2x^2_0 \right)\left( h-m-\frac{m}{2} B^2{x_0}^2 +B\intop_1^1S_1(h-m)dx_1\right) dx_0}                        \nonumber \\
&\simeq&  B{\displaystyle\intop_{-1}^{1} \left( h-m +2BS_1(h-m)\right)dx_0} =  2(h-m)B + 4(h-m)S_1B^2
\end{eqnarray}
the density of thermodynamics, up to the same order, is
\begin{eqnarray}
j&=&-\frac{\partial}{\partial h}f(h,m)= \frac{m}{2\pi}\intop_{-B}^{B}d\vartheta \varepsilon\left( \vartheta\right)\cosh \vartheta = \frac{m}{2\pi}B\intop_1^1\left( 1 + S_1B\intop_1^1dy\right) dx= \nonumber  \\
&=&\frac{m}{\pi}\left(B+ 2S_1B^2\right) 
\end{eqnarray}
so that the energy density becomes
\begin{eqnarray}
\Omega&=&\frac{m}{2\pi}B\intop_1^1\left( 1 + S_1B\intop_1^1dy\right)-\frac{m}{2\pi} \left( 2(h-m)B + 4(h-m)S_1B^2\right)+O\left(B^3\right)=\nonumber  \\ 
&=&\frac{m^2}{\pi} \left(B+2S_1B^2\right) +O\left(B^3\right) = mj + O(j^3)
\end{eqnarray}
Indeed, contibutions of order $B^2$ are all included in the $mj$ term. Unfortunately, to catch the $j^3$ contribution we have to keep the $B^3$s, as well as to catch the $j^n$ contribution we have to keep the $B^n$s, an consider at least (n-1) nested kernels in the pseudoenergy.

%%%%%%%%%%%%%%%%%%%%%%%%%%%%%%%%%%%%%%%%%%%%%%%%%%%%%%%%%%%%%%%%%%%%%%%%%%%%%%%%%%%%%%%%%%%%%%%%%%%%%%%%%%%%%%%%%%%%%%%

\subsection{What we can do up to the order $B^3$}
Once again, we briefly repeat the very same steps as before, reaching one higher power of $B$.
\begin{equation}
\varepsilon \left(\vartheta\right)=
h-m\cosh\vartheta+B\intop_{-1}^{1}dxK\left(\vartheta-Bx\right)\left[h-m\cosh\left(Bx\right)+B\intop_{-1}^{1}dyK\left(Bx-By\right)\left(h-m\right)\right]
\end{equation}
\begin{eqnarray}
f &\simeq & -\frac{m}{2\pi}B\intop_{-1}^{1}dx\left(1-\frac{1}{2}B^{2}x^{2}\right) \nonumber \\
 & &\qquad \qquad\left\{h-m-\frac{m}{2}B^{2}x^{2}+B\intop_{-1}^{1}dxK\left(Bx-By\right)
   \left[h-m-\frac{m}{2}B^{2}x^{2}+2B\frac{1}{\pi}S_{1}\left(h-m\right)\right]\right\}   \nonumber \\
& \simeq & -\frac{m}{2\pi}B\intop_{-1}^{1}dx\left(1-\frac{1}{2}B^{2}x^{2}\right)\left[h-m-\frac{m}{2}B^{2}x^{2}+2B\frac{1}{\pi}S_{1}\left(h-m\right)+4B^{2}\left(\frac{1}{\pi}S_{1}\right)^{2}\left(h-m\right)\right]
\end{eqnarray}
\begin{eqnarray}
j&=&-\frac{\partial f}{\partial h}\simeq\frac{m}{2\pi}\intop_{-B}^{B}d\vartheta\cosh\vartheta\left[1+2B\frac{1}{\pi}S_{1}+4B^{2}\left(\frac{1}{\pi}S_{1}\right)^{2}\right]  \nonumber \\
&\simeq & \frac{m}{\pi}\left(B+\frac{1}{6}B^{3}\right)\left[1+2B\frac{1}{\pi}S_{1}+4B^{2}\left(\frac{1}{\pi}S_{1}\right)^{2}\right]= \nonumber \\
&\simeq &\frac{m}{\pi}\left(B+2\frac{1}{\pi}S_{1}B^{2}+4\frac{1}{\pi^{2}}S_{1}^{2}B^{3}+\frac{1}{6}B^{3}\right)
\end{eqnarray}
\begin{eqnarray}
\Omega&=&f+jh\simeq\frac{m^{2}}{2\pi}B\intop_{-1}^{1}dx\left(1-\frac{1}{2}B^{2}x^{2}\right)\left[1+\frac{1}{2}B^{2}x^{2}+2B\frac{1}{\pi}S_{1}+4B^{2}\left(\frac{1}{\pi}S_{1}\right)^{2}\right]= \nonumber \\
&=&mj+\frac{m^{2}}{6\pi}B^{3}
\end{eqnarray}
now, the term with $B^3$ can come only from the third power of the density $j$, because we have already acknowledged the term linear in $j$, so we calculate the coefficient of the $j$-expansion of $\Omega$ as the ratio among $\frac{m^{2}}{6\pi}$ and the coefficient of the first power of $B$ in the expansion of $j$:$\qquad \Omega_{3}=\frac{\frac{m^{2}}{6\pi}}{\left(\frac{m}{\pi}\right)^{3}}=\frac{\pi^{2}}{6m}$. So we have
\begin{equation}
\Omega = mj + \frac{\pi^2}{6m}j^3 +O(j^4)
\end{equation}

%%%%%%%%%%%%%%%%%%%%%%%%%%%%%%%%%%%%%%%%%%%%%%%%%%%%%%%%%%%%%%%%%%%%%%%%%%%%%%%%%%%%%%%%%%%%%%%%%%%%%%%%%%%%%%%%%%%%%%%%%%%%%%%%%
\subsection{A small improvement}
We saw in (\ref{B2})that $B\approx \sqrt{2\frac{h-m}{m}}+O\left(\frac{h-m}{m}\right)$. We can use this to show the stability of the non-interacting fermion gas approximation and that the most crude approximation yelds a correct coefficient of $j^3$.
\begin{eqnarray}
\varepsilon\left(\vartheta\right)&=&h-m\cosh\vartheta+
\intop_{-B}^{B}d\vartheta'K\left(\vartheta-\vartheta'\right)\left[h-m\cosh\left(\vartheta'\right)  +B\intop_{-1}^{1}d\vartheta''K\left(\vartheta'-\vartheta''\right)\left(h-m\right)\right]+O(B^{3}) \nonumber \\
&=&h-m\cosh\vartheta+B\intop_{-1}^{1}d\vartheta'\frac{1}{\pi}S_{1}\left[h-m\cosh\left(\vartheta'\right)+2B\frac{1}{\pi}S_{1}\left(h-m\right)\right]+O(B^{3})= \nonumber \\
&=&h-m\cosh\vartheta+B\frac{1}{\pi}S_{1}\intop_{-1}^{1}dx\left[h-m\cosh\left(Bx\right)+BS_{1}\intop_{-1}^{1}dy\left(h-m\right)\right]+O(B^{3})= \nonumber \\
&=&h-m\cosh\vartheta+R(B)
\end{eqnarray}
where $R(B)=2\left(hB-m\sinh B\right) +4 B^2\left(h-m\right) \simeq 2(h-m)(B+2B^2)$ behaves as a modification of the chemical potential. Thus, we can expect that our result will not be different from the free one. Indeed
\begin{eqnarray}
f(h) & = & -\frac{m}{2\pi}\intop_{-B}^{B}d\vartheta\left(h+R\left(B\right)-m\cosh\vartheta\right)\cosh\vartheta \nonumber  \\
& = & \frac{m}{2\pi}\intop_{-B}^{B}d\vartheta\left(h+2(h-m)(B+2B^{2})-m\left(1-\frac{\vartheta^{2}}{2}B^{2}\right)\right)\left(1-\frac{\vartheta^{2}}{2}B^{2}\right)+O(B^{4})= \nonumber \\
& = & -\frac{m}{\pi}\left((h-m)B+2(h-m)(B+2B^{2})\right)+O(B^{4})
\end{eqnarray}
We easily calculate, from (\ref{h-derivative}),
\begin{equation}
\frac{\partial}{\partial h}\varepsilon\left(\vartheta\right)=1+2S_{1}B+4S_{1}B^{2}+O(B^{3})
\end{equation}
and, from (\ref{density j})
\begin{equation}
j=-\frac{\partial}{\partial h}f=\frac{m}{2\pi}\intop_{-B}^{B}\frac{\partial}{\partial h}\varepsilon\left(\vartheta\right)\cosh\vartheta d\vartheta=\frac{m}{\pi}B\left(1+2S_{1}B+4S_{1}B^{2}\right)+O(B^{4})
\end{equation}
to conclude from (\ref{legendre}), somehow more "efficiently", that 
\begin{equation} \label{Omegarapida}
\Omega=jm-\frac{m}{\pi}\left((h-m)B+2(h-m)(B+2B^{2})+\left(h-m\right)2B\left(1+2S_{1}B+4S_{1}B^{2}\right)\right)+O(B^{4})
\end{equation}
now, by remembering that $ h-m \simeq mB^2+O(B^3)$, it becomes evident how the structure of the forcing therm ensures the stability of the free-fermion approximation: every correction affects the series from the $B^4$ (i.e. from the $j^4$) term on. Moreover, due to the absence of the $B$s and of the $B^2$s, it suffices the first-order approximation for $B$ to fix the coefficient of $j^3$.

%%%%%%%%%%%%%%%%%%%%%%%%%%%%%%%%%%%%%%%%%%%%%%%%%%%%%%%%%%%%%%%%%%%%%%%%%%%%%%%%%%%%%%%%%%%%%%%%%%%%%%%%%%%%%%%%%%%%%%%%%%%%%%%%%%

\section{A method for calculating $B$ to all orders}
We have seen that we can express $B$ in powers of $x=\frac{h-m}{m}$ as
\begin{equation}\label{B(x)}
B=\sum_{n=0}^\infty b_n x^n
\end{equation}
as well as we can write the pseudoenergy as a power series
\begin{equation}\label{e(B)}
\varepsilon (B)=m\sum_{n=0}^\infty e_n B^n
\end{equation}
At the same time, the x-dependence of the coefficients $e_n$ can be explicited by Taylor expansion. This means that
\begin{eqnarray}
\varepsilon\left( B(x)\right) &=&e_0^{(0)}+e_0^{(1)}x+e_0^{(2)}x^2+\ldots+\left(e_1^{(0)}+e_1^{(1)}x+e_1^{(2)}x^2+\ldots\right)\left(b_0+b_1x+b_2x^2+\ldots\right)+  \nonumber\\ 
&&+\left(e_2^{(0)}+e_2^{(1)}x+e_2^{(2)}x^2+\ldots\right)\left(b_0+b_1x+b_2x^2+\ldots\right)^2+\ldots
\end{eqnarray}
We saw that, up to the second order, the only nonzero coefficients were
\begin{displaymath} \label{b coefficients}
 \begin{array}{cccc}
 e_0^{(2)}=1 & e_1^{(2)}=\frac{2}{\pi}S_1 & e_2^{(0)}=-\frac{1}{2} & e_2^{(2)} = \frac{4}{\pi^2}S_1^2
\end{array}
\end{displaymath}
so we can solve, order by order, the boundary condition (\ref{bound})
\begin{displaymath}
 \begin{array}{cccc} 
   zero & -\frac{1}{2}b_0=0 & \Rightarrow & b_0=0  \nonumber\\
   one & 0=0 & & useless                          \nonumber\\
   two & e_0^{(2)}+e_1^{(1)}b_1+e_1^{(0)}b_2+e_2^{(0)}b_1^2+b_0 (\ldots )=1-\frac{1}{2}b_1^2=0 & \Rightarrow & b_1=\sqrt{2}   \nonumber\\
   three & e_1^{(2)}b_1+2e_2^{(0)}b_1b_2+e_3^{(0)}b_1^3 +b_0 (\ldots ) & \Rightarrow & b_2=2\frac{S_1}{\pi}+2e_3^{(0)}   \nonumber\\
   \dots & \ldots&  &
\end{array}
\end{displaymath}
the zeroth order fixes $b_0$: this guarantees that it is possible to calculate $b_k$ only with the first $k$ orders in $B$ of (\ref{bound}). In general
\begin{equation}  \label{bienne}
b_{n-1}=-\frac{1}{2e_2^{(0)}b_1}\sum_{m=1}^n\sum_{t=0,2}\left(1-\delta_{m,2}\delta_{t,0}\right) e_m^{(t)}\sum_{j_1+\ldots+j_n=n-t}b_{j_1}\ldots b_{j_m}
\end{equation}
with
\begin{eqnarray}\label{e coefficients}
e_n^{(0)}&=&\sum_{p=0}^\infty\sum_{n+2p+2k_1+\ldots +2k_n=N} \frac{(-1)^{k_1+\ldots +k_n}}{\pi^n(2p)!}
           \intop_{-1}^1 dx_1 \ldots \intop_{-1}^1 dx_n \left(1-x_1\right)^{2k_1} \ldots \left(x_{n-1}-x_n \right)^{2k_n} x_n^{2p}B^{N}
           \nonumber  \\
e_n^{(2)}&=&\sum_{n+2k_1+\ldots +2k_n=N}\frac{(-1)^{k_1+\cdots +k_n}}{\pi^n}
           \intop_{-1}^1 dx_1 \ldots \intop_{-1}^1 dx_n \left(1-x_1\right)^{2k_1} \ldots \left(x_{n-1}-x_n \right)^{2k_n} B^{N}
\end{eqnarray}
%%%%%%%%%%%%%%%%%%%%%%%%%%%%%%%%%%%%%%%%%%%%%%%%%%%%%%%%%%%%%%%%%%%%%%%%%%%%%%%%%%%%%%%%%%%%%%%%%%%%%%%%%%%%%%%%%%%%%%%

\section{$\Omega(j)$: a systematics to all orders}
By substituting the explicit formula for the kernel in (\ref{solution})
\begin{eqnarray} \label{kernel}
K\left( \vartheta-\vartheta'\right)&=& {\displaystyle \frac{1}{\pi^{2}} \sum_{k=0}^{\infty}}\left[\frac{1}{k+1}\frac{\left(-1\right)^{k}}{1+\left(\frac{\vartheta-\vartheta'}{\pi(k+1)}\right)^{2}}+\frac{1}{k+2\Delta}\frac{\left(-1\right)^{k}}{1+\left(\frac{\vartheta-\vartheta'}{\pi(k+2\Delta)}\right)^{2}}\right]=  \nonumber \\
&=&\frac{1}{\pi^{2}}{\displaystyle \sum_{k=0}^{\infty}}\left[\frac{\left(-1\right)^{k}}{k+1}{\displaystyle \sum_{j=0}^{\infty}\left(-1\right)^{j}\left(\frac{\vartheta-\vartheta'}{\pi(k+1)}\right)^{2j}}+\frac{\left(-1\right)^{k}}{k+2\Delta}{\displaystyle \sum_{j=0}^{\infty}\left(-1\right)}\left(\frac{\vartheta-\vartheta'}{\pi(k+2\Delta)}\right)^{2j}\right] \nonumber \\
\end{eqnarray}
we get
\begin{eqnarray}
\varepsilon(\vartheta)&=& h-m\cosh\vartheta+  \nonumber  \\
&+&{\displaystyle\sum_{n=1}^{\infty}\frac{1}{\pi^{2n}}B^{n}}\intop_{-1}^{1}dx_{1}...\intop_{-1}^{1}dx_{n}{\displaystyle \sum_{m_{1}=0}^{\infty}} \left[\frac{1}{k+1}\frac{\left(-1\right)^{k}}{1+\left(\frac{\vartheta-Bx_{1}}{\pi(k+1)}\right)^{2}}+\frac{1}{k+2\Delta}\frac{\left(-1\right)^{k}}{1+\left(\frac{\vartheta-Bx_{1}}{\pi(k+2\Delta)}\right)^{2}}\right]\ldots        \nonumber \\ 
&&\ldots{\displaystyle\sum_{m_{n}=1}^{\infty}}\left[\frac{1}{m_{n}+1}\frac{\left(-1\right)^{k}}{1+\left(\frac{x_{n-1}-x_{n}}{\pi(k+1)}\right)^{2}}+\frac{1}{m_{n}+2\Delta}\frac{\left(-1\right)^{k}}{1+\left(\frac{x_{n-1}-x_{n}}{\pi(k+2\Delta)}\right)^{2}}\right]\left(h-m\cosh\left(Bx_{n}\right)\right)=  \nonumber  \\
&=&h-m\cosh\vartheta+{\displaystyle\sum_{n=1}^{\infty}\frac{1}{\pi^{2n}}B^{n}}\intop_{-1}^{1}dx_{1}...\intop_{-1}^{1}dx_{n}{\displaystyle\sum_{m_{1}=0}^{\infty}}{\displaystyle\left(-1\right)^{m_{1}}\sum_{j_{1}=0}^{\infty}\left(-1\right)^{j_{1}}}   \nonumber  \\
& &\qquad\left[\frac{1}{m_{1}+1}{\displaystyle \left(\frac{\vartheta-\vartheta'}{\pi(m_{1}+1)}\right)^{2j_{1}}}+\frac{1}{m_{1}+2\Delta}\left(\frac{\vartheta-\vartheta'}{\pi(m_{1}+2\Delta)}\right)^{2j_{1}}\right] {\displaystyle \sum_{m_{1}=0}^{\infty}}{\displaystyle\left(-1\right)^{m_{1}}\sum_{j_{1}=0}^{\infty}\left(-1\right)^{j_{1}}}...           \nonumber     \\
&&\qquad\left[\frac{1}{m_{1}+1}{\displaystyle\left(\frac{\vartheta-\vartheta'}{\pi(m_{1}+1)}\right)^{2j_{1}}}+\frac{1}{m_{1}+2\Delta}\left(\frac{\vartheta-\vartheta'}{\pi(m_{1}+2\Delta)}\right)^{2j_{1}}\right]\left(h-m\cosh\left(Bx_{n}\right)\right)
\end{eqnarray}
and by defyining the series
\begin{displaymath}
 S_{n}={\displaystyle \sum_{k=0}^{\infty}}\left[\frac{\left(-1\right)^{k}}{\pi^{n}\left(k+1\right)^{n}}+\frac{\left(-1\right)^{k}}{\pi^{n}\left(k+2\Delta\right)^{n}}\right]
\end{displaymath}
we can rewrite the previous result in a much more tangled way, but with a more transparent insight on the contributions of the different powers of $B$ 
\begin{eqnarray}
\varepsilon(\vartheta) &=& h-m\cosh\vartheta+  \nonumber  \\
&&{\displaystyle \sum_{n=1}^{\infty}\frac{B^{n}}{\pi^{n}}}\intop_{-1}^{1}dx_{1}...\intop_{-1}^{1}dx_{n}\sum_{j_{1}...j_{n}=0}^{\infty}\left(-1\right)^{j_{1}+...+j_{n}}S_{j_{1}}{\displaystyle B^{j_{1}}\left(\frac{\vartheta}{B}-Bx_{1}\right)^{2j_{1}}} 
S_{j_{2}}B^{j_{2}}{\displaystyle \left(x_{1}-x_{2}\right)^{2j_{2}}}\ldots \nonumber \\  &&\qquad\qquad\qquad\qquad\qquad\qquad\qquad\ldots S_{j_{n}}B^{j_{n}}\left(x_{n-1}-x_{n}\right)^{2j_{n}}\left(h-m\cosh\left(Bx_{n}\right)\right)
\end{eqnarray}
Please note that we are allowed to exchange the order integrations and summations over the $j_i$ indexes, because on the finite interval $\left[-1,1\right]$ all powers of $x$ are integrable and the succession of reduced sum does converge to the actual kernel.
\begin{eqnarray}
\varepsilon\left( \vartheta\right) &=&  h-m\cosh\vartheta+\nonumber \\
&+&{\displaystyle\sum_{n=1}^{\infty}\frac{1}{\pi^{n}}}\sum_{j_{1}...j_{n}=0}^{\infty}\left(-1\right)^{J_{n}}B^{n+2J_{n}}S_{2j_{1}+1}S_{2j_{2}+1}...S_{2j_{n}+1}{\displaystyle \intop_{-1}^{1}dx_{1}...\intop_{-1}^{1}dx_{n}\left(\frac{\vartheta}{B}-Bx_{1}\right)^{2j_{1}}}... \nonumber \\
&&\qquad\qquad\qquad\qquad\qquad\qquad\dots{\displaystyle\left(x_{n-1}-x_{n}\right)^{2j_{n}}}\left(h-m{\displaystyle\sum_{p=0}^{\infty}}B^{b}\frac{x_{n}^{2p}}{\left(2p\right)!}\right)=   \nonumber\\
&=& h-m\cosh\vartheta+ \nonumber  \\
&+&{\displaystyle \sum_{n=1}^{\infty}}\sum_{j_{1}...j_{n}=0}^{\infty}\frac{\left(-1\right)^{J_{n}}}{\pi^{n-J_{n}}}S_{2j_{1}+1}S_{2j_{2}+1}...S_{2j_{n}+1}{\displaystyle \intop_{-1}^{1}dx_{1}...\intop_{-1}^{1}dx_{n}\left(\frac{\vartheta}{B}-Bx_{1}\right)^{2j_{1}}}{\displaystyle \left(x_{1}-x_{2}\right)^{2j_{2}}}...\nonumber\\
&&\qquad\qquad\qquad\qquad\qquad\qquad\ldots{\displaystyle\left(x_{n-1}-x_{n}\right)^{2j_{n}}}\left(h-m{\displaystyle \sum_{p=0}^{\infty}}B^{b}\frac{x_{n}^{2p}}{\left(2p\right)!}\right)B^{n+2J_{n}}
\end{eqnarray}
where $ J_{n}=j_{1}+j_{2}+...+j_{n}$. In the following we will also define
\begin{equation}
S_n\left( \left\lbrace j \right\rbrace \right)=S_{2j_{1}+1}S_{2j_{2}+1}...S_{2j_{n}+1}
\end{equation}
following \cite{HMN} we can compute the free energy as a function of the chemical potential $h$ and of the mass $m$ of the field
\begin{eqnarray}\label{complete free energy}
-\frac{2\pi}{m} f\left( h,m\right) = {\displaystyle\intop_{-B}^{B} \cosh\vartheta \varepsilon \left( \vartheta\right)d\vartheta} =B{\displaystyle\intop_{-1}^{1} \cosh\left( B\,x_0 \right)  \varepsilon \left( B\,x_0 \right)dx_0} = \nonumber \\
= \Bigg\{h{\displaystyle \sum_{r=0}^{\infty}\frac{1}{(2r)!} \sum_{n=0}^{\infty}\frac{1}{\pi^n} \sum_{j_1 j_2 \ldots j_n\;=0}^{\infty} \left( -1\right) ^{J_n} S_n\left( \left\lbrace j \right\rbrace \right)} \intop_{-1}^{1}dx_0 ... \intop_{-1}^{1}dx_n     \nonumber \\
\qquad\qquad\qquad\ldots x_0^{2r}\left( x_0-x_1\right) ^{2j_1}\ldots \left( x_{n-1}-x_n\right) ^{2j_1} B^{1+n+2J_n+2r} + \nonumber\\
-m{\displaystyle \sum_{r=0}^{\infty}\frac{1}{(2r)!} \sum_{n=0}^{\infty}\frac{1}{\pi^n} \sum_{j_1 j_2 \ldots j_n\;=0}^{\infty} \left( -1\right) ^{J_n} S_n\left( \left\lbrace j \right\rbrace \right) \sum_{p=0}^{\infty}\frac{1}{(2p)!}} \intop_{-1}^{1}dx_0 ... \intop_{-1}^{1}dx_n     \nonumber \\
\qquad\ldots x_0^{2r}\left( x_0-x_1\right) ^{2j_1}\ldots \left( x_{n-1}-x_n\right) ^{2j_1} x_n^{2p} B^{1+n+2J_n+2r+2p}\Bigg\} 
\end{eqnarray}
now it is to the density $j$, dual to the chemical potential $h$ through the Helmholtz free energy $f$.
\begin{eqnarray}
j&=&-\frac{\partial}{\partial h}f(h,m)={\displaystyle \frac{m}{2\pi}\frac{\partial}{\partial h}\intop_{-B}^{B}d\vartheta \varepsilon\left( \vartheta\right)\cosh \vartheta = \frac{m}{2\pi}\intop_{-B}^{B}d\vartheta \left(1+ \intop_{-B}^{B}d\vartheta' K \left( \vartheta-\vartheta'\right) \frac{\partial\varepsilon_h(\vartheta')}{\partial h} \right) \cosh\vartheta  } \nonumber \\
&=&\frac{m}{2\pi}{\displaystyle \sum_{r=0}^{\infty}\frac{1}{(2r)!} \sum_{n=0}^{\infty}\frac{1}{\pi^n} \sum_{j_1 j_2 \ldots j_n\;=0}^{\infty} \left( -1\right) ^{J_n} S_n\left( \left\lbrace j \right\rbrace \right)} \intop_{-1}^{1}dx_0 ... \intop_{-1}^{1}dx_n   \nonumber \\
&&\qquad\qquad\qquad\ldots x_0^{2r}\left( x_0-x_1\right) ^{2j_1}\ldots \left( x_{n-1}-x_n\right) ^{2j_1} B^{1+n+2J_n+2r}
\end{eqnarray}
We have used (\ref{hderivative}) and (\ref{bound}). The Legendre transform follows straightforwardly
\begin{eqnarray}
\Omega\left( j,m \right)&=&f\left( h,m\right) + jh= \nonumber  \\
&=&\frac{m^2}{2\pi}{\displaystyle \sum_{r=0}^{\infty}\frac{1}{(2r)!} \sum_{n=0}^{\infty}\frac{1}{\pi^n} \sum_{j_1 j_2 \ldots j_n\;=0}^{\infty} \left( -1\right) ^{J_n} S_n\left( \left\lbrace j \right\rbrace \right) \sum_{p=0}^{\infty}\frac{1}{(2p)!}} \intop_{-1}^{1}dx_0 ... \intop_{-1}^{1}dx_n     \nonumber \\
&&\qquad\qquad\qquad x_0^{2r}\left( x_0-x_1\right) ^{2j_1}\cdots \left( x_{n-1}-x_n\right) ^{2j_1} x_n^{2p} B^{1+n+2J_n+2r+2p}= \nonumber  \\
&=&mj+\frac{m^2}{2\pi}{\displaystyle \sum_{r=0}^{\infty}\frac{1}{(2r)!} \sum_{n=0}^{\infty}\frac{1}{\pi^n} \sum_{j_1 j_2 \cdots j_n\;=0}^{\infty} \left( -1\right) ^{J_n} S_n\left( \left\lbrace j \right\rbrace \right) \sum_{p=1}^{\infty}\frac{1}{(2p)!}} \intop_{-1}^{1}dx_0 ... \intop_{-1}^{1}dx_n     \nonumber \\
&&\qquad\qquad\qquad x_0^{2r}\left( x_0-x_1\right) ^{2j_1}\cdots \left( x_{n-1}-x_n\right) ^{2j_1} x_n^{2p} B^{1+n+2J_n+2r+2p}
\end{eqnarray}
Please note that the $p$ index in the last summation, having recognized the $p=0$ term as the linear contribution in $j$, now runs from $1$ to infinity. This, of course, allows us to exclude any term of order $j^2$, as the lowest power of $B$ in the summation is now three. At this step, we have the two series
\begin{eqnarray}
 \Omega(B)&=&{\displaystyle \sum_{n=1}^\infty \omega _n B^n} \\
 j(B)&=&{\displaystyle \sum_{n=1}^\infty j_n B^n}
\end{eqnarray}
In this situation, one would naturally appeal to Lagrange inversion formula and express $\Omega$ in powers of $j$ as
\begin{equation} \label{Energia}
 \Omega(j)={\displaystyle \sum_{n=1}^\infty \Omega _n j^n}={\displaystyle \sum_{n=1}^\infty \frac{1}{n!}\left[ \frac{d^{n-1}}{dB^{n-1}}\left( \phi ^n(B) \Omega'(B) \right)\right]_{B=0}   j^n}
\end{equation}
where $\phi(B)=\frac{B}{j(B)}$.

\subsection{Some explicit coefficient}
However, in order to perform some calculation, it may be simpler to extract the coefficient of the $n$-th power of $j$, $\Omega _n$, from the coefficient of the powers of $B$ up to the $n$-th: $\omega _1,\quad\ldots,\quad\omega _n$. In fact, having already isolated the term $mj$, we easily calculate $\Omega _3$:
\begin{equation}
 \omega _1=\frac{m}{2\pi}\intop_{-1}^1dx_0=\frac{m}{\pi}
\end{equation}
\begin{equation}
 \omega _3=\frac{m^2}{2\pi}\frac{1}{2}\intop_{-1}^1{x_0}^2dx_0=\frac{m^2}{6\pi}
\end{equation}
\begin{equation}
 \Omega _3=\frac{\omega _3}{{j_1}^3}=\frac{\pi^2}{6m}
\end{equation}
We can make another step without too much trouble
\begin{equation}
 \omega _4=\frac{m^2}{2\pi}\frac{1}{2}\frac{S_1}{\pi} \intop_{-1}^1dx_0 \intop_{-1}^1dx_1 {x_1}^2=\frac{m^2}{6\pi^2} = \frac{m^2}{3\pi^2}S_1
\end{equation}
\begin{equation}
 \left( j^3\right) _4=3\frac{m}{\pi}\frac{m}{\pi}\frac{m}{2\pi}\frac{S_1}{\pi} \intop_{-1}^1dx_0 \intop_{-1}^1dx_1 =3\frac{2m^3}{\pi^4}S_1 
\end{equation}
we will now specialize the solution to the O(6) sigma model, i.e. we set $N=6$, so that $\Delta=\frac{1}{4}$
\begin{equation} \label{S_1}
 S_1={\displaystyle \frac{1}{\pi}\sum_{k=0}^\infty\left[\frac{(-1)^k}{k+1}+\frac{(-1)^k}{k+\frac{1}{2}}\right]} = \frac{1}{\pi}\ln 2 + \frac{1}{2}
\end{equation}
\begin{equation}
 \Omega _4=\frac{\omega _4-\Omega _3(j^3)_4}{{j_1}^4}=\frac{1}{\frac{m^4}{\pi^4}}\left[ \frac{m^2}{3\pi^2}S_1- \frac{\pi^2}{6m}\frac{6m^3}{\pi^4}S_1\right] = -\frac{2}{3} \frac{\pi^2}{m^2}\left( \frac{1}{\pi}\ln 2 + \frac{1}{2} \right)
\end{equation}
where we have written the coefficient of $j^4$ in the series in $B$ of $j^3$ as $(j^3)_4$. Now we have warmed up, we can approach the next one:
\begin{eqnarray}
 \omega _5&=& \frac{m^2}{\pi}\left(\frac{7}{120}+\frac{2}{3}\frac{1}{\pi^2}S_1^2\right)   \nonumber \\
\left(j^{4}\right)_{5}&=&4j_{1}^{3}j_{2}=4\left(\frac{m}{\pi}\right)^{3}\frac{m}{2\pi}\frac{8}{\pi}S_{1}  \nonumber \\
\left(j^{3}\right)_{5}&=&3j_{1}^{2}j_{3}+3j_{1}j_{2}^{2}=\left(\frac{m}{\pi}\right)^{3}\left(\frac{12}{\pi^{2}}S_{1}^{2}+\frac{1}{2}+\frac{12}{\pi^{2}}S_{1}^{2}\right)
\end{eqnarray}
\begin{eqnarray}
\Omega_{5}&=&\frac{1}{j_{1}^{5}}\left(\omega_{5}-\Omega_{4}\left(j^{4}\right)_{5}-\Omega_{3}\left(j^{3}\right)_{5}\right)=\frac{1}{j_{1}^{5}}\left(\omega_{5}-3\Omega_{3}j_{1}^{2}j_{3}-3\Omega_{3}j_{1}j_{2}^{2}-4\Omega_{4}j_{1}^{3}j_{2}\right)=  \nonumber \\
&=& \frac{\pi^4}{m^3}\left(-\frac{1}{40}+\frac{2}{\pi^2}S_1^2\right)
\end{eqnarray}

On one hand, please note that the coefficient $\Omega_5$, as well as those of all odd powers of $j$, does reduce to its free approximation if we switch off the interaction, i.e., if we send $S_1 \rightarrow 0$. On the other hand, since the oddness of the non-interacting gas series, the $\Omega_4$  correcly vanishes in this limit. Generally speaking, unfortunately, when considering the order $B^n$, we need to subtract all the terms coming from lower powers of $j$, which makes the calculation quite cumbersome.

%%%%%%%%%%%%%%%%%%%%%%%%%%%%%%%%%%%%%%%%%%%%%%%%%%%%%%%%%%%%%%%%%%%%%%%%%%%%%%%%%%%%%%%%%%%%%%%%%%%%%%%%%%%%%%%%%%%%%
\subsection{Comparison with Lagrange formula}
Of course, this results does match with those coming from Lagrange formula. The first coefficient is trivial, the second reads
\begin{eqnarray}
 \Omega_2 &=& \frac{1}{2}\frac{d}{dB}\left[ \frac{{\displaystyle\sum_{p=1}^\infty p\omega _pB^{p-1}}} {\left({\displaystyle\sum_{k=1}^\infty j_k B^{k-1}}\right) ^2}\right]_{B=0}=
\frac{1}{2}\left[ \frac{{\displaystyle\sum_{p=2}^\infty} p(p-1)\omega _pB^{p-2}}{\left( {\displaystyle\sum_{k=1}^\infty} j_k B^{k-1}\right) ^2} - 2\frac{ {\displaystyle\sum_{p=1}^\infty p\omega _pB^{p-1} \sum_{k=2}^\infty \left( k-1\right) j_k B^{k-2} }}{\left( {\displaystyle\sum_{k=1}^\infty} j_k B^{k-1}\right) ^3}\right]_{B=0} =  \nonumber  \\
&=&\frac{1}{2}\left( \frac{2\omega_2}{j_1}-2\frac{j_2}{j_1}\right) =0
\end{eqnarray}
Checks. The third
\begin{eqnarray}
 \Omega_3 &=& \frac{1}{6}\left[ \Omega'''\Phi^3 + 6 \Omega''\Phi^2 \Phi' + 3\Omega'\left( \Phi^2 \Phi''+\Phi\left( \Phi'\right)^2 \right) \right] _{B=0} \nonumber\\
\end{eqnarray}
\begin{eqnarray}
 \Omega'(0) &=& \omega _1  \nonumber  \\
 \Omega''(0) &=& 2\omega _2   \nonumber  \\
  &\cdots&   \nonumber  \\
 \Omega^{\left( n\right) }(0) &=& n! \omega _n 
\end{eqnarray}
\begin{eqnarray}
 \Phi'(0) &=& \frac{d}{dB}\frac{B}{J(B)}_{B=0}=-\left[ \frac{\sum_{p=2}^\infty (p-1)j_pB^{p-2}}{\left( \frac{j}{b}\right) ^2} \right] _{B=0}  = - \frac{j_2}{{j_1}^2}              \nonumber  \\
 \Phi''(0) &=& 2\frac{{j_2}^2-j_1j_3}{{j_1}^3}
\end{eqnarray}
\begin{eqnarray}
 \Omega_3 &=& \frac{1}{6}\left[ 6\frac{\omega _3 }{{j_1}^3} - 12 \frac{\omega_2 j_2}{{j_1}^4} + 3\omega_1 
 \left( 2\frac{{j_2}^2}{{j_1}^5}+2\frac{{j_2}^2-j_1j_3}{{j_1}^5}\right) \right] _{B=0}                 \nonumber\\
&=&\frac{1}{6}\left[6\frac{\omega _3 -j_3}{j_1^3} -12\frac{{j_2}^2}{j_1^4}+12\frac{\omega_1}{j_1}\frac{{j_2}^2}{j_1^4} \right] =		\nonumber\\ 
&=&\frac{1}{6}\frac{\pi^4}{m^4}6\left(\frac{m}{2\pi}\frac{2}{3}+\frac{m}{2\pi} \frac{8}{\pi^2} S_1- \frac{m}{2\pi}\frac{1}{3}-
   \frac{m}{2\pi}\frac{8}{\pi^2}S_1\right) = \frac{\pi^2}{6m}
\end{eqnarray}
checks again. We will not go further.

%%%%%%%%%%%%%%%%%%%%%%%%%%%%%%%%%%%%%%%%%%%%%%%%%%%%%%%%%%%%%%%%%%%%%%%%%%%%%%%%%%%%%%%%%%%%%%%%%%%%%%%%%%%%%%%%%%%%%%%%%%%%%%%%%%%%%%%%%%%%%%%%%
\subsection{Analiticity}
The convergence of the two series in powers of $B$ will be estimated by combinatorial considerations. We have $k^N$ possibilities to arrange $k$ nonnegative integers in such a way that their sum yelds $N$. Moreover, the number $n$ of the factors appears in the exponent together with the factors $j_n$ themselves and the two indices $p$, $s$. So the general $N$-th coefficient grows less than
\begin{displaymath}
\sum_{n=0}^{N}2^{2n}\left(N-n\right)^{n+2}
\end{displaymath}
where we have used
\begin{displaymath}
\intop_{-1}^1dx_0\ldots \intop_{-1}^1dx_n \cdots\left(x_{0}-x_{1}\right)^{2j_{1}}\ldots\left(x_{n-1}-x_{n}\right)^{2j_{n}} \ll 2^n\cdot2^n
\end{displaymath}
and
\begin{displaymath}
|S_{n}|=|{\displaystyle \sum_{k=0}^{\infty}}\left[\frac{\left(-1\right)^{k}}{\pi^{n}\left(k+1\right)^{n}}+\frac{\left(-1\right)^{k}}{\pi^{n}\left(k+2\Delta\right)^{n}}\right]| \leq 1
\end{displaymath}
We find the maximum value of the addend 
\begin{eqnarray} \frac{d}{dx}\left(N-x\right)^{x+2}&=&e^{\left(x+2\right)\ln\left(N-x\right)}\left(\ln\left(N-x\right)-\frac{x+2}{N-x}\right) = 0\qquad\Longrightarrow\qquad\left(N-a\right)=e^{\frac{a+2}{N-a}} \nonumber\\
\frac{d^{2}}{dx^{2}}\left(N-x\right)^{x+2}&=&\frac{d}{dx}e^{\left(x+2\right)\ln\left(N-x\right)}\left(\ln\left(N-x\right)-\frac{x+2}{N-x}\right)=  \nonumber  \\
&=&e^{\left(x+2\right)\ln\left(N-x\right)}\left[\left(\ln\left(N-x\right)-\frac{x+2}{N-x}\right)^{2}-\left(\frac{2}{N-x}+\frac{x+2}{\left(N-x\right)^{2}}\right)\right]= \nonumber  \\
&=&-e^{\left(a+2\right)\ln\left(N-a\right)}\ln\left(N-a\right) \left(\frac{2}{N-a}+\frac{a+2}{\left(N-x\right)^{2}}\right)  <0  \qquad\textrm{(in $x=a)$}
\end{eqnarray}
and we substitute it in the sum
\begin{equation}
{\displaystyle \sum_{n=0}^{N}2^{2n}\left(N-n\right)^{n+2}}\leq 2^{2N}{\displaystyle \sum_{n=0}^{N}e^{\frac{a+2}{N-a}a+2}}=2^{2N}\left(N+1\right)e^{\frac{\left(a+2\right)^{2}}{N-a}}\leq 2^{2N} \left(N+1 \right) e^{\frac{\left(N+2\right)^{2}}{N}}\sim N\left( 4e\right) ^{N}
\end{equation}
The general term grows with the $N$-th power, so that the two series of powers of $B\ll 1$ easily converge. To conclude, Lagrange theorem expresses the analiticity of the series (\ref{Energia}), when written in powers of $\frac{j}{m}\ll 1$.

\section{Checks and previsions for strong SYM$_4$}
As noticed by \cite{AM}, the strong coupling limit $j<<g$ and the consequent reduction to the O(6) bosonic nonlinear sigma model in two dimensions allows to calculate exactly the the anomalous dimension of high spin operators $f(g,j)$ with $j=\frac{J}{\log S}$: the SO(6) charge density modifies the scaling by a $2\Omega$, 
\begin{equation} \label{adden}
f(g,j)=f_0(g)+2\Omega(g,j)
\end{equation}
The possibility of calculating, at least in principle, many successive coefficient $f_1, f_2, \ldots$ of the expansion (as announced in the footnote $10$ of \cite{FGR1}) allows a check of the first scaling functions from the SYM side, appearing in the Sudakov scaling (\ref{Suds}). At this purpose, we underline that, on the one hand, the density $j$ that we have used is defined from the string side in (3.1) of \cite{AM} as
\begin{equation}
j=\frac{J}{2\log S}
\end{equation}
while on the other hand, recently available calculations \cite{FGR2} approach the problem with slightly different notations. From their point of view
\begin{equation}
j_{SYM}=\frac{J}{\log S}=2j
\end{equation}
It follows that, rescaling their results, we get an explicit link with the O(6) model, namely
\begin{equation}
f_n=2^{n-1}\Omega_n
\end{equation}
The first three coefficients from the sigma model point of view ($\Omega_1$, $\Omega_2$ and $\Omega_3$), are exactly the same as for the free, non-relativistic theory. They were already given as an approximation in \cite{HMN}, as we said above. Further, we have performed the calculation of $\Omega_4$ and $\Omega_5$, that contains trace of our interaction and the model-dependence, too, entering through $S_1$ as in (\ref{S_1}).
On the other front, the correspondence with $f_1$, $f_2$, $f_3$ and $f_4$ is a remarkable fact. In particular, this very last scaling function catches the nature of the system and selects the specific model.

\medskip

{\bf Acknowledgments} D.F. owe very much to Marco Rossi and Paolo Grinza for 
many suggestions, and thanks the INFN grant "Iniziativa specifica PI14" and the
international agreement INFN-MEC-2008 for travel financial support. The authors also thank 
D. Bombardelli, F. Ravanini for useful discussions.

\end{document}